\documentclass[12pt,a4paper]{article}
\usepackage[latin1]{inputenc}
\usepackage[T1]{fontenc}
\usepackage{amsmath}
\usepackage{amsfonts}
\usepackage{amssymb}
\usepackage[english]{babel}
\usepackage[dvips]{graphicx,color}
\title{New tests and applications of the worldline path integral in
  the first order formalism }
\author{C.~D.~Fosco$^{a}$\,,
 J.~S\'anchez-Guill\'en$^{b}$ and
  R.~A.~V\'azquez$^{b}$\\
  {\normalsize\it $^a$Centro Atómico Bariloche and Instituto Balseiro}\\
  {\normalsize\it Comisión Nacional de Energ\'{\i}a Atómica}
  \\
  {\normalsize\it 8400 Bariloche, Argentina.}\\
  {\normalsize\it $^b$Departamento de F\'{\i}sica de  Part\'{\i}culas}\\
  {\normalsize\it Facultad de F\'{\i}sica and Instituto Galego de Altas
Energ\'{\i}as} \\
  {\normalsize\it Universidad de Santiago}\\
  {\normalsize\it E-15782 Santiago de Compostela, Spain.} }
\begin{document}
%
\date{}
\maketitle
\begin{abstract}
\noindent We present different non-perturbative calculations within 
the context of Migdal's representation for the propagator and
effective action of quantum particles. We first calculate the exact
propagators and effective actions for Dirac, scalar and Proca fields
in the presence of constant electromagnetic fields, for an
even-dimensional spacetime.  Then we derive the propagator for a
charged scalar field in a spacelike vortex (i.e., instanton)
background, in a long-distance expansion, and the exact propagator for
a massless Dirac field in $1+1$ dimensions in an arbitrary background.
Finally, we present an interpretation of the chiral anomaly in the
present context, finding a condition that the paths must fulfil in
order to have a non-vanishing anomaly.
\end{abstract}
\section{Introduction}\label{sec:intro}
Worldline formulations have been applied since a long time
ago~\cite{schwinger} to the derivation of many interesting Quantum
Field Theory results. More recent applications have emerged as a
by-product of the new insights gained by the re-derivation of
worldline representations by taking the infinite tension limit in
(perturbative) string theory amplitudes~\cite{Bern:1992ad}, and by the
introduction of new ways to handle the spin degrees of
freedom~\cite{Strassler:1992zr,Karanikas:1995ua,Karanikas:1995zi}.
Besides, elegant proposals to treat more general situations, involving
internal degrees of freedom and general couplings to higher-spin
fields have been advanced~\cite{D'Hoker:1995ax,D'Hoker:1995bj}.

In these methods, different sets of variables and alternative
constructions have been used in order to `exponentiate' the relevant
observables and then perform the path integral.  In spite of the {\em
formal\/} equivalence between the different methods, there are few
concrete calculations that may serve as tests to gain a deeper
understanding of the method and about the physics involved.  Important
steps in that direction have already been taken; indeed, some
non-perturbative calculations corresponding to external constant
electromagnetic fields have been obtained within the worldline
representation~\cite{Reuter:1996zm}. Another setting were the
worldline approach can be independently tested is in numerical
calculations~\cite{nos}.

In this article, we present new tests corresponding to concrete
examples, obtained within the worldline path-integral representation
for Dirac fields in the first order formalism introduced by Migdal
in~\cite{migdal} and further extended
in~\cite{Karanikas:1995ua,Karanikas:1995zi}. The first order formalism
preserves the geometrical picture and is quite intuitive (for example,
it does not involve Grassmann variables). These two features are, we
believe, among the main advantages of the worldline method. It is also
more adequate for some numerical computations which are specially
suited for non perturbative calculations.

\noindent  We shall follow our previous work~\cite{Fosco}, where some 
features of the method have been discussed in detail, including a
proof of the equivalence with standard Quantum Field Theory at the
perturbative level.

The structure of this paper is as follows: in section~\ref{sec:meth} we
briefly review the main properties of the representation introduced
in~\cite{migdal}, with emphasis in the objects we shall be concerned with
in the examples. To elucidate the quite general nature of this approach, we
also introduce the worldline representation for the propagator and the
effective action corresponding to a Proca field.

In~\ref{sec:fdet} we deal with a constant $F_{\mu\nu}$ field in $1+1$
dimensions, for the Dirac, scalar and Proca cases.  In
section~\ref{sec:dqt2}, we generalize the previous cases to $d > 2$ (
$d \,=\,{\rm even}$) dimensions.

Section~\ref{sec:vortex} contains a derivation of the scalar
propagator in a vortex-like background in $1+1$ dimensions, and
section~\ref{sec:exact} the calculation of the exact propagator for a
massless Dirac field in $1+1$ dimensions.

In section~\ref{sec:anomaly} we present a study of the kind of
`roughness' one should expect the trajectories to have, in order to
contribute to the chiral anomaly. We show that there has to be a
singularity in the correlation between different time derivatives of
the paths, whose precise form depends on the number of spacetime
dimension.

Finally, section~\ref{sec:concl} presents our conclusions.

\section{The method}\label{sec:meth}
Our aim here is to calculate the propagator and effective action
corresponding to external electromagnetic fields.  We shall be
concerned with Dirac, scalar and Proca models, coupled to Abelian
gauge fields.

Let us consider first the case of a massive Dirac field in $d$ Euclidean
dimensions, whose action, $S_f$, has the following form:
\begin{equation}\label{eq:defsf}
S_f({\bar\psi},\psi,A) \;=\; \int d^dx \, {\bar\psi} (\not\!\! D +
m) \psi \;,
\end{equation}
where
\begin{equation}\label{eq:dirconv}
\not\!\! D \equiv \gamma_\mu D_\mu \;\;,
\;\;\; D_\mu = \partial_\mu + i e A_\mu \;\;,
\;\;\; \gamma_\mu^\dagger
= \gamma_\mu \;\;,\;\;\;\mu = 1,\ldots,d \;.
\end{equation}
The $\gamma_\mu$ matrices satisfy the Clifford Algebra:
\begin{equation}
\{\gamma_\mu , \gamma_\nu\} = 2 \, \delta_{\mu\nu}\;\;\;\; \forall \mu =
1,\ldots,d \;,
\end{equation}
$A_\mu$ denotes an Abelian gauge field and $e$ is a coupling constant with the
dimensions of $[{\rm mass}]^{\frac{4-d}{2}}$.

In the worldline formulation of~\cite{migdal}, the fermion propagator,
denoted here by $G_f(x,y)$, is represented by the path integral:
\begin{eqnarray}\label{eq:dprop}
G_f(x,y) &=& \int_0^\infty dT \,e ^{- m T} \, \int_{x(0) = y}^{x(T) = x}
{\mathcal D}p
{\mathcal D}x \; e^{i \int_0^T d\tau  p(\tau) \cdot{\dot x}(\tau) } \nonumber\\
&×& {\mathcal P}[ e^{- i \int_0^T d\tau {\not p}(\tau) }] \;\; e^{- i e
\int_0^T d\tau {\dot
x}(\tau)\cdot A[x(\tau)]} \;,
\end{eqnarray}
where we have explicitly indicated the boundary conditions for the
$x_\mu(\tau)$ paths.  The $p_\mu(\tau)$ paths are, on the other hand,
unconstrained.

Another object we will be interested in is $\Gamma_f(A)$, the (normalized)
contribution of the fermionic determinant to the effective action:
\begin{equation}\label{eq:defga}
\Gamma_f(A) \;\equiv\; - \ln \Big[ \frac{\det \big(\not \!\! D + m \big)}{\det
  \big(\not \! \partial + m \big)} \Big]
\;=\; - {\rm Tr} \ln (\not \!\! D + m \big) \;+\;
{\rm Tr} \ln  (\not \! \partial + m \big) \;,
\end{equation}
which (by definition) verifies $\Gamma_f(0)=0$.

For $\Gamma_f(A)$ we have the worldline representation:
$$
\Gamma_f (A) \;=\; \int_{0}^\infty \frac{dT}{T} \, e^{- m T} \,
\int_{x(0)=x(T)} {\mathcal D}p \,  {\mathcal D}x \; e^{ i \int_0^T d\tau
 p_\mu(\tau) {\dot x}_\mu(\tau) } $$
\begin{equation}\label{eq:gppa}
× \; {\rm tr}\big[ {\mathcal P} e^{-i \int_0^T d\tau \not p(\tau)} \big]
\; e^{-i e \int_0^T d\tau {\dot x}_\mu(\tau) A_\mu[x(\tau)] }\;,
\end{equation}
where the functional integration measure may be formally represented as:
\begin{equation}\label{eq:measure}
{\mathcal D}p {\mathcal D}x \;\equiv\; \prod_{0 < \tau \leq T}
\frac{d^dx(\tau) d^dp(\tau)}{(2 \pi)^d}\;,
\end{equation}
and it is (also formally) dimensionless, since there are as many $dp$'s as
there are $dx$'s in the integration measure and, in our conventions, $\hbar
=1$. Of course, this formal definition can be made more rigorous by
introducing a discrete approximation to it, and taking the corresponding
limit. This procedure will, indeed, be used later on to deal with some
examples.

We will also consider complex scalar fields, their propagators (to be denoted
by $G_b$) and their contribution to the effective action ($\Gamma_b$). In the
context of the worldline formulation we have explained before, those objects
have similar expressions to their Dirac counterparts. Indeed, if the field
theory action $S_b$ for $\varphi,\,{\bar \varphi}$ is:
\begin{equation}
S_b \;=\; \int d^dx \, \big[ {\bar D_\mu \varphi} D_\mu \varphi \, + \, m^2
{\bar \varphi} \varphi \big] \;,
\end{equation}
then, an entirely analogous definition to the one used for the Dirac field
leads to
\begin{eqnarray}\label{eq:dbprop}
G_b(x,y) &=& \int_0^\infty dT \, e^{- m^2 T}\,  \int_{x(0) = y}^{x(T) = x}
{\mathcal D}p
{\mathcal D}x \; e^{i \int_0^T d\tau p(\tau)\cdot{\dot x}(\tau)} \nonumber\\
&×& e^{- \int_0^T d\tau \, p^2(\tau) } \;\; e^{- i e \int_0^T d\tau {\dot
x}(\tau)\cdot A[x(\tau)]} \;,
\end{eqnarray}
and
$$
\Gamma_b (A) \;=\; - \int_{0}^\infty \frac{dT}{T} \, e^{- m^2 T} \,
\int_{x(0)=x(T)}
{\mathcal D}x {\mathcal D}p \; e^{ i \int_0^T d\tau p_\mu(\tau) {\dot
x}_\mu(\tau) }
$$
\begin{equation}\label{eq:gbppa}
× \; e^{- \int_0^T d\tau \, p^2(\tau)} \; e^{-i e \int_0^T d\tau {\dot
x}_\mu(\tau) A_\mu[x(\tau)] }\;.
\end{equation}

It is interesting to compare the previous expressions with their Dirac
field counterparts: note that the difference amounts to replacing the
object
\begin{equation}
\Phi_f (T) \;\equiv\; {\mathcal P} \big[ e^{- i \int_0^T d \tau \,{\not
p}(\tau) } \big],
\end{equation}
by
\begin{equation}
\Phi_b (T) \;\equiv\; e^{- \int_0^T d \tau \,p^2(\tau) },
\end{equation}
in the corresponding fermionic formula. Besides, there is a $(-1)$
factor in $\Gamma_b$ because of the different statistics; $m$ is replaced by
$m^2$, and the trace of $\Phi_b$ is of course absent.

This general structure will reproduce itself with more or less
straightforward changes for the next example that we shall consider:
the Proca field, for which we use $a_\mu$ to denote the field variable,
to avoid confusion with the gauge field $A_\mu$.  The Euclidean action,
$S_P$, is defined by:
\begin{equation}
S_{P} \;=\; \int d^dx \, \Big( \frac{1}{4} f_{\mu\nu} f_{\mu\nu}
+ \frac{1}{2} m^2 a_\mu a_\mu   \Big) \;.
\end{equation}
where $f_{\mu\nu} = \partial_\mu a_\nu - \partial_\nu a_\mu$.  This action corresponds of course
to the case of a {\em real\/} field, for which it makes sense to define
the (free) propagator, but to allow for a coupling to an external
gauge field $A_\mu$, we also consider the complex field version:
\begin{equation}
S_{P}(a^*,a;A) \;=\; \int d^dx \, \big( \frac{1}{2}
|D_\mu a_\nu - D_\nu a_\mu |^2 +  m^2 |a|^2   \big)
\end{equation}
with the covariant derivative $D_\mu = \partial_\mu + i e A_\mu$.

Based on the form of the Euclidean actions, it is rather
straightforward to derive the propagator in the presence of an
external field $A_\mu$. Indeed, we have
\begin{eqnarray}\label{eq:dPprop}
G_P(x,y) &=& \int_0^\infty dT \, e^{- m^2 T}\,  \int_{x(0) = y}^{x(T) = x}
{\mathcal D}p
{\mathcal D}x \; e^{i \int_0^T d\tau p(\tau)\cdot{\dot x}(\tau)} \nonumber\\
&×&  \Phi_P (T)  \;\; e^{- i e \int_0^T d\tau {\dot
x}(\tau)\cdot A[x(\tau)]} \;,
\end{eqnarray}
with
\begin{equation}
\Phi_P (T) \;=\; {\mathcal P} \exp{\Big[- \int_0^T d\tau p_\alpha(\tau)
  p_\beta(\tau) \Gamma^P_{\alpha \beta}} \Big]
\end{equation}
and the $\Gamma^P_{\alpha\beta}$ are a set of $d × d$ matrices whose
components are:
 \begin{equation}
\big[\Gamma^P_{\alpha \beta} \big]_{\mu\nu} \;=\; \delta_{\alpha\beta}
\delta_{\mu\nu} - \frac{1}{2} \big( \delta_{\alpha\mu} \delta_{\beta\nu}
+ \delta_{\alpha\nu} \delta_{\beta\mu} \big) \;.
\end{equation}

It should now be clear that, when considering the one loop effective
action one needs to evaluate the expression:
\begin{eqnarray}\label{eq:dPloop}
\Gamma_P (A) &=& - \int_0^\infty \frac{dT}{T} \, e^{- m^2 T}\,
\int_{x(0) = x(T)}
{\mathcal D}p {\mathcal D}x \; e^{i \int_0^T d\tau p(\tau)\cdot{\dot x}(\tau)} \nonumber\\
&×& {\rm tr} \big[ \Phi_P (T) \big]  \;\; e^{- i e \int_0^T d\tau {\dot
x}(\tau)\cdot A[x(\tau)]} \;.
\end{eqnarray}

A quite remarkable property, that we may use to our advantage, is that the
functional integral over $x_\mu$, for a given background field $A_\mu$ is the
same for all the fields. The differences shall of course appear when
evaluating the integrals over $p_\mu$, since they are affected by the
spin-dependent factor $\Phi (T)$.

We conclude this section by mentioning that the previous representations are
not unique (in many ways). One of the reasons is that one may always describe
a theory (with any spin) in terms of first order equations, although for a
different set of field variables. Indeed, the equations of motion for a free
field $\varphi$ may always be written as follows~\cite{korch}:
\begin{equation}
\big( \Gamma_\mu \partial_\mu + m \big) \phi(x) \;=\; 0 
\end{equation}
where $\phi$ is a multicomponent field, defined in terms of $\varphi$ and its
derivatives, while $\Gamma_\mu$ are matrices whose form depend on the spin
content of the original field $\varphi$. For example, for a massive real
scalar field $\varphi$, we may write the action:
\begin{equation}
S \;=\, \frac{m}{2} 
\int d^dx \, \bar{\phi} ( \Gamma_\mu \partial_\mu + m) \phi \;,
\end{equation}
where the $\Gamma_\mu$ are, in this case, the $(d+1) × (d+1)$ matrices
\begin{equation}
\big[ \Gamma_\mu \big]_{ab}\;=\; \delta_{a,\mu + 1} \delta_{\nu + 1, b}
\end{equation}
and 
\begin{equation}
\phi \;=\; 
 \left( 
       \begin{array}{c} 
         \varphi \\
         -\partial_1\varphi / m\\
         -\partial_2\varphi / m\\
          \ldots\\
         - \partial_d\varphi /m
       \end{array}
 \right)\;.
\end{equation}
The `adjoint' $\bar{\phi}$ is defined as: $\bar{\phi} = \phi^T \Gamma_0$, 
 where
\begin{equation}
 \Gamma_0  \;=\; 
 \left( 
  \begin{array}{cc} 
        1     &   0_{1× d} \\  
        0_{d × 1}   &   I_{d× d}  
  \end{array}
 \right) \;.
\end{equation}
If the field is instead complex, and it is coupled to an external gauge field
$A_\mu$, we have
\begin{equation}
S \;=\, m \, \int d^dx \, \bar{\phi} ( \Gamma_\mu D_\mu + m) \phi \;,
\end{equation}
with the only difference with respect to the previous case in that 
$\bar{\phi} = \phi^\dagger \Gamma_0$. 

The generalization to higher spins $J$ is simple, although care must be taken
when considering $J >1$~\cite{korch}, due to the existence of non-trivial
constraints on the state vectors, depending on representation chosen for the
$\Gamma_\mu$ matrices.

Once this first-order formulation is introduced, one may write a world-line
representation for the one-loop effective action $\Gamma$, which is given by:
$$
\Gamma (A) \;=\; \int_{0}^\infty \frac{dT}{T} \, e^{- m T} \,
\int_{x(0)=x(T)} {\mathcal D}p \,  {\mathcal D}x \; e^{ i \int_0^T d\tau
 p_\mu(\tau) {\dot x}_\mu(\tau) } $$
\begin{equation}
× \; {\rm tr}\big[ {\mathcal P} e^{-i \int_0^T d\tau \Gamma_\mu p_\mu (\tau)} 
\big] \; e^{-i e \int_0^T d\tau {\dot x}_\mu(\tau) A_\mu[x(\tau)] }\;,
\end{equation}
which has the same structure as the one introduced for the Dirac
case~\footnote{The structure is more complicated for $J > 1$,
  see~\cite{korch}.}.
\section{Constant external field in $1+1$ dimensions}\label{sec:fdet}
\subsection{Dirac field}
We shall present here the evaluation of the fermionic determinant and
propagator for a massive Dirac field in the presence of a constant
external $F_{\mu\nu}$ field, in $1+1$ dimensions.  As usual, rather
than working directly with the determinant, we instead use the
effective action $\Gamma_f(A)$,
$$
\Gamma_f(A) \,=\, \int_0^\infty \frac{dT}{T} \, e^{- m T} \,
\int_{x(0)=x(T)} {\mathcal
  D}p \,{\mathcal D} x \;\; e^{i \int_0^T d\tau p_\mu (\tau) {\dot x}_\mu
(\tau)}
$$
\begin{equation}
× \; {\rm tr} \Big[ {\mathcal P} e^{- i \int_0^T d\tau \not p (\tau)}\Big]
e^{-i e
\int_0^T d\tau \, {\dot x}_\mu(\tau) A_\mu(x(\tau)) },
\end{equation}
where $A_\mu$ is such that:
\begin{equation}
F_{\mu\nu} \;=\; \partial_\mu A_\nu - \partial_\nu A_\mu \;=\; F \,
\varepsilon_{\mu\nu}\;,
\end{equation}
with $F \equiv {\rm constant}$. For $A_\mu$ we adopt a gauge-fixing condition
such that
\begin{equation}
A_1 (x) \;=\, - F\, x_2 \;\;,\;\;\;\; A_2 \;=\; 0\;.
\end{equation}
We then see that:
\begin{equation}\label{eq:add1}
\Gamma_f(A) \,=\, \int_0^\infty \frac{dT}{T} \, e^{- m T} \, \int
{\mathcal D}p \;\;{\rm tr}\Big[ {\mathcal P} e^{- i \int_0^T d\tau
\not p (\tau)}\Big] \; {\mathcal Z}(p,F)
\end{equation}
with
\begin{equation}\label{eq:add2}
{\mathcal Z}(p,F)\;=\; \int_{x(0)=x(T)}\,  {\mathcal D}x \;\; e^{i
\int_0^T d\tau \, p_\mu (\tau) {\dot x}_\mu (\tau)} \;  e^{i e F \int_0^T
d\tau \, {\dot x}_1 (\tau) x_2(\tau) }\;.
\end{equation}
We shall now evaluate ${\mathcal Z}(p,F)$. As it will become clear,
the same object appears within the context of the complex scalar field
determinant.

To evaluate, we first separate it into two iterated integrals, one for each
component:
$$
{\mathcal Z}(p,F)\;=\; \int_{x_2(0)=x_2(T)}\,  {\mathcal D}x_2 \;\Big\{
e^{i \int_0^T d\tau {\dot x}_2 (\tau) \, p_2(\tau)}  \;
$$
\begin{equation}
× \; \int_{x_1(0)=x_1(T)}\,  {\mathcal D}x_1 \; e^{i \int_0^T d\tau {\dot
x}_1 (\tau)
[ e F x_2(\tau) \,+\, p_1(\tau)]} \Big\}\;.
\end{equation}
The two previous integrals are quite simple to evaluate, the result
being:
\begin{equation}
{\mathcal Z}(p,F)\;\propto\; \exp[ - \frac{i}{e F} \int_0^T d\tau \, {\dot
p}_1(\tau)
p_2(\tau) ] \;,
\end{equation}
an expression which captures the exact dependence on $p_\mu(\tau)$.
However, in order to calculate $\Gamma_f(A)$ exactly, we need to know the
exact form of ${\mathcal Z}(A)$, including any relevant global factor.
One safe way to do that is, as usual, to introduce a discretization of
the functional integral. For example, splitting the $[0,T]$ interval
into $n$ sub-intervals, we see that the functional integral over
$x_1(\tau)$ is given by the limit:
$$
\int_{x_1(0)=x_1(T)}\,  {\mathcal D}x_1 \; e^{i \int_0^T d\tau {\dot x}_1
(\tau)
[ e F x_2(\tau) \,+\, p_1(\tau)]}
$$
\begin{equation}
= \; \lim_{n \to \infty} \left\{ \int \Big( \prod_{k=1}^n dx_1^{(k)} \Big)
\;
e^{i \sum_{k=1}^n ( x_1^{(k+1)} - x_1^{(k)} ) [ e F x_2^{(k)} \,+\,
  p_1^{(k)}]} \right\} \;,
\end{equation}
where $x_1^{(k)}$ denotes $x_1(\tau)$ at the discrete time $\tau_k$, with
$\tau_k \equiv \frac{k T}{n}$, and a similar convention for $x_2$ and $p_\mu$.
Periodicity requires $x_\mu^{(n+1)} = x_\mu^{(1)}$. It is then immediate
to see that:
$$
 \int \Big( \prod_{k=1}^n dx_1^{(k)} \Big) \; e^{i \sum_{k=1}^n (
x_1^{(k+1)} - x_1^{(k)} ) [ e F x_2^{(k)} \,+\,
  p_1^{(k)}]}
$$
\begin{equation}
\;=\; L_1 \, \prod_{k=1}^{n-1} 2 \pi \, \delta\Big( e F (x_2^{(l)} -
x_2^{(l-1)})
\,+\, p_1^{(l)} - p_1^{(l-1)} \Big) \;,
\end{equation}
where $L_1$ is the total length of the system along the $x_1$
coordinate.  Discretizing also the $x_2(\tau)$ integral, an analogous
calculation yields:
\begin{eqnarray}
{\mathcal Z}(p,F) &=& L_1 \, L_2 \; \lim_{n\to\infty}  \;\left\{ (\frac{2
\pi}{e
  F})^{n-1}\, e^{- \frac{i}{e F} \,  \sum_{k=1}^n p_2^{(k)} (
    p_1^{(k+1)} - p_1^{(k)} )} \right\} \nonumber\\
&=& \frac{ e F L_1 L_2}{2 \pi}  \; \lim_{n\to\infty}  \; \left\{  (\frac{2
\pi}{e
  F})^{n}\, e^{- \frac{i}{e F} \,  \sum_{k=1}^n p_2^{(k)} (
    p_1^{(k+1)} - p_1^{(k)} )} \right\} \nonumber\\
&=& \xi  \; \lim_{n\to\infty}  \; \left\{ (\frac{2 \pi}{e
  F})^{n}\, e^{- \frac{i}{e F} \,  \sum_{k=1}^n p_2^{(k)} (
    p_1^{(k+1)} - p_1^{(k)} )} \right\} \;,
\end{eqnarray}
where $L_2$ is the system length along the second coordinate. We have
factored-out the dimensionless quantity \mbox{$\xi \equiv \frac{ e F L_1
    L_2}{2 \pi}$}, which measures the `flux' through the system's area
$L_1 L_2$, in units of the elementary flux.

Note also that the product of $\delta$ functions implies, in particular,
that the integral over $p_1(\tau)$ (to be performed next), will be over a
space of periodic paths. Namely, the integral over $x$ enforces
periodic boundary conditions for the integral over $p_1$.

Then we insert the previous result for ${\mathcal Z}(p,T)$ into the
expression for $\Gamma_f(A)$, and see that:
$$
\Gamma_f(A) \,=\, \xi \, \int_0^\infty  \frac{dT}{T} \, e^{- m T} \,
\int_{p_1(0)=p_1(T)} \widehat{{\mathcal D}p} 
$$
\begin{equation}
× \, {\rm tr}\Big[ {\mathcal P} e^{- i \int_0^T d\tau \not p (\tau) } \Big]
\; e^{- \frac{i}{e F} \, \int_0^T d\tau \, {\dot p}_1(\tau) p_2(\tau)},
\end{equation}
where the new integration measure for $p(\tau)$, $\widehat{{\mathcal
    D}p}$, is defined by:
\begin{equation}
\widehat{{\mathcal D}p}\;=\; \prod_{0 < \tau \leq T} \frac{d p_1(\tau)
dp_2(\tau)}{2
  \pi e F},
\end{equation}
(note that one of the two $(2\pi)$ factors from (\ref{eq:measure})
cancels out). In the integral over $p(\tau)$, due to the presence of the
term $\int d\tau {\dot p}_1(\tau) p_2(\tau)$ in the exponent, the functional
integral is equivalent to the operatorial trace of an evolution
operator, with the $p_\mu$'s replaced by time-independent, noncommuting
operators:
\begin{equation}
\Gamma_f (A) \;=\; \xi \,  \int_0^\infty \frac{dT}{T} \, e^{- m T}
\; {\rm Tr} \Big( e^{- i \, T \,  \not {\hat p} } \Big),
\end{equation}
where the ${\hat p}_\mu$'s satisfy the commutation relation:
\begin{equation}
[ {\hat p}_1 \,,\,{\hat p}_2 ] \;=\;  - i e F \;,
\end{equation}
and the trace is over Hilbert and Dirac spaces.

To evaluate that trace we first write the operator $\not \! {\hat p}$
more explicitly, as follows:
\begin{equation}
\not \! {\hat p} \;=\; \sqrt{2 e F} \; {\widehat{\mathcal O}},
\end{equation}
where
\begin{equation}
{\widehat{\mathcal O}}\,=\, \left( \begin{array}{cc}
0 & {\hat a} \\ {\hat a}^\dagger & 0 \end{array} \right),
\end{equation}
and ${\hat a} \equiv \frac{{\hat p}_1 - i {\hat p}_2}{\sqrt{2 e F}}$,
${\hat a}^\dagger \equiv \frac{{\hat p}_1 + i {\hat p}_2}{\sqrt{2 e F}}$ (we
assume that $e F >0$).

Since the operators ${\hat a}$ and ${\hat a}^\dagger$ verify $[{\hat a},
{\hat a}^\dagger]=1$, we can calculate the spectrum of the self-adjoint
operator ${\widehat {\mathcal O}}$ exactly. Indeed, we find the exact
eigenvalues and eigenvectors to be the following:
\begin{eqnarray}\label{eq:eigen}
{\widehat{\mathcal O}} | \varphi_n^{(±)} \rangle &=& \lambda_n^{(±)} \, |
\varphi_n^{(±)} \rangle
\;,\;\; n \in {\mathbb N} \nonumber\\
{\widehat{\mathcal O}} |\varphi_0\rangle &=& 0 \;,
\end{eqnarray}
where
\begin{equation}
\lambda_n^{(±)} \;=\;± \, \sqrt{n} \;, \;\;\; n \;=\; 1,\, 2,\; \ldots
\end{equation}
and
\begin{eqnarray}
|\varphi_n^{(±)} \rangle &=& \frac{1}{\sqrt{2}} \; \left(
\begin{array}{c}
± | n-1 \rangle \\
| n \rangle
\end{array}
\right) \;\;\; n \;=\;1,\,2,\, \ldots \nonumber\\
|\varphi_0\rangle &=& \left( \begin{array}{c}
0 \\
| 0 \rangle
\end{array}
\right) \;.
\end{eqnarray}
Here, $|n\rangle$ denotes the (normalized) eigenstates of the `number'
operator ${\hat a}^\dagger {\hat a}$. Note that the upper element in $|\varphi_0\rangle$
is $0$ (the null vector), while the lower one is the `vacuum' state.

Then the effective action becomes:
\begin{equation}
 \Gamma_f(A) \;=\; \xi \;  \int_0^\infty \frac{dT}{T} \,
e^{- m T} \Big[1\;+\; \sum_{n=1}^\infty  \big( e^{- i T \sqrt{2 e F n}
}\;+\; e^{i
  T \sqrt{2 e F n}} \big) \Big],
\end{equation}
or, integrating out $T$:
\begin{equation}
 \Gamma_f(A) \;=\; \xi \; \Big[ \ln m \,+\,  \sum_{n=1}^\infty \ln ( m^2 +
2 e F
 n)  \Big] \;,
\end{equation}
where we have neglected a constant which is independent of $F$ and $m$.

Now to sum up the series, we use the representation:
\begin{equation}
\ln x \;=\; -  \lim_{s\to0} \, \frac{d}{ds} [ x^{- s} ],
\end{equation}
to obtain:
\begin{equation}
 \Gamma_f(A) \;=\;  \xi  \; \lim_{s \to 0}
   \frac{d}{ds} \;\left\{  \sum_{n=1}^\infty (m^2 + 2 e F n)^{-s}   +
m^{-s}  \right\} \;,
\end{equation}
or:
\begin{equation}
 \Gamma_f(A) \;=\; \xi  \; \lim_{s \to 0} \frac{d}{ds} \; \left\{
\sum_{n=1}^\infty (2 e F)^{-s} ( n  + \frac{m^2}{2 e
       F})^{-s}   + m^{-s} \right\} \;,
\end{equation}
and
\begin{equation}
\Gamma_f(A) \;=\; \xi  \; \lim_{s \to 0} \frac{d}{ds} \; \left\{ (2 e
F)^{-s} \zeta_H(s; 1 + \frac{m^2}{2 e F}) + m^{-s} \right\} \;,
\end{equation}
where $\zeta_H$ denotes Hurwitz $\zeta$-function. The effective action is then
obtained by taking the limit explicitly, and it is coincident with the
results of~\cite{Blau:1988iz}, namely:
$$
\Gamma_f(A) \;=\; L_1 L_2  \;\Big[  \frac{ e F + m^2}{4\pi} \ln ( 2 e F)
$$
\begin{equation}
+\, \frac{e F}{2 \pi} \ln \Gamma( 1 +\frac{m^2}{2 e F}) - \frac{e F}{4 \pi}
\ln (2
\pi m^2) \Big] \;.
\end{equation}
 The imaginary part of ${\tilde
  \Gamma}_{(1+1)}(A)$ in Minkowski spacetime may be obtained by Wick
rotating: $F \to i F$, so that:
\begin{equation}
\Im [{\tilde \Gamma}_{(1+1)}(A)] \;=\; \sum_{n=1}^\infty \,
{\rm arctan}[\frac{2 e F n}{m^2}]\;,
\end{equation}
which can be also written in terms of the dimensionless vacuum
angle~\cite{Coleman:1976uz} $\theta$ for the massive Schwinger model
\begin{equation}
\theta = \frac{2 \pi F}{e},
\end{equation}
as:
\begin{equation}
\Im [{\tilde \Gamma}_{(1+1)}(A)] \;=\; \sum_{n=1}^\infty \,
{\rm arctan}[(\frac{e^2}{m^2})\; \frac{\theta  n}{\pi}]\;.
\end{equation}
The result for the imaginary part does not exhibit the periodicity in
$\theta$ of the interacting model, since here the gauge field is not
dynamical.

The procedure we have followed for the calculation of the effective
action may of course also be applied to the propagator, if one takes
into account the main differences: namely, that the integration over
$x$ is not over periodic paths, and that the spin degrees of freedom
are not traced.
 Thus we are lead to:
\begin{equation}
G_f(x,y) \;=\; \int_0^\infty \, dT \, e^{- m T}
\; \langle x | e^{- i \, T \,  \not {\hat p} } | y \rangle \,,
\end{equation}
with the same definition for ${\hat p}$ we had in the effective action
calculation. In abstract operatorial form:
\begin{equation}
G_f \;=\; \int_0^\infty \, dT \, e^{- m T}
\; e^{- i \, T \,  \not {\hat p} }\,,
\end{equation}
and matrix elements may be taken with respect to any convenient
basis. Since we already know the eigenvectors of ${\hat p}$, we can
use that basis. Integrating out over the `time' $T$ the result is
\begin{equation}
G_f \,=\, \frac{1}{m} {\mathcal P}_0 \,+\,
\sum_{n=1}^\infty \Big\{ \frac{2 m}{m^2 + 2 e F n} {\mathcal P}_n
\,+\, \frac{-2 i m}{m^2  + 2 e F n} {\mathcal Q}_n \Big\}
\end{equation}
where
\begin{equation}
{\mathcal P}_0 \;=\; \left( \begin{array}{cc}
0 & 0 \\ 0 & |0\rangle \langle 0 | \end{array}
\right)
\end{equation}
\begin{equation}
{\mathcal P}_n \;=\; \left( \begin{array}{cc}
|n-1\rangle \langle n-1| & 0 \\ 0 & |n\rangle \langle n|\end{array}
\right) \;\;\; (\forall n > 1)\;,
\end{equation}
and
\begin{equation}
{\mathcal Q}_n \;=\; \left( \begin{array}{cc} 0 & |n-1\rangle \langle
 n| \\
|n\rangle \langle n-1| & 0 \end{array} \right) \;\;\; (\forall n > 1)\;.
\end{equation}
\subsection{Complex scalar field}
Let us now consider the changes that arise when calculating the
effective action $\Gamma_b$, for the same gauge field configuration.
First, we note that the calculation of ${\mathcal Z}(p,F)$ goes
through in the same way as for the Dirac case, and we directly arrive
to
\begin{equation}
\Gamma_b (A) \;=\; - \xi \; \int_0^\infty \frac{dT}{T} \, e^{- m^2 T} \;
{\rm Tr}
\Big[ e^{- \, T \,   {\hat p}_\mu {\hat p}_\mu  }\Big],
\end{equation}
where the ${\hat p}_\mu$ operators are the same as the ones from the
Dirac field calculation. The trace ${\rm Tr}$ is now over the Hilbert
space only.  In terms of the destruction and creation operators ${\hat
  a}$, ${\hat a}^\dagger$ used in the previous subsection, we see that:
\begin{equation}
\Gamma_b (A) \;=\; - \xi \; \int_0^\infty \frac{dT}{T} \, e^{- m^2 T} \;
{\rm Tr}
\Big[ e^{- \, T \,  2 e F ( {\hat{\mathcal N}} + \frac{1}{2})  }\Big],
\end{equation}
where ${\hat{\mathcal N}}$ is the number operator corresponding to
${\hat a}$ and ${\hat a}^\dagger$:
\begin{equation}
 {\hat{\mathcal N}} \;=\; {\hat a}^\dagger {\hat a}\;.
\end{equation}
Then we write the trace in terms of the eigenvalues,
\begin{equation}
\Gamma_b (A) \;=\; - \xi \; \int_0^\infty \frac{dT}{T} \, e^{- m^2 T} \;
\sum_{n=0}^\infty e^{- \, T \,  2 e F ( n + \frac{1}{2}) },
\end{equation}
and integrate over $T$ to obtain:
\begin{equation}
\Gamma_b (A) \;=\; - \xi \;\sum_{n=0}^\infty \ln \big[ m^2 \,+\,  e F ( 2 n
+ 1)
\big]\;.
\end{equation}
Of course, this may be evaluated as in the Dirac case in terms of
Hurwitz $\zeta$ function:
\begin{equation}
\Gamma_b (A) \;=\;  \xi  \frac{d}{ds} \Big[ ( 2 e F)^{-s} \, \zeta_H\big(s;
\frac{1}{2} +
\frac{m^2}{2 e F}\big) \Big] {\Big |}_{s=0} \;.
\end{equation}
Again, the result is identical to the one of~\cite{Blau:1988iz}.

The scalar propagator, $G_b$ is simpler than its Dirac
counterpart. Indeed, a straightforward calculation yields:
\begin{equation}
G_b \;=\; \int_0^\infty \, dT \, e^{- m^2 T}
\; e^{- \, T \,2 e F \, ( {\hat a}^\dagger {\hat a} + \frac{1}{2} )  }\,,
\end{equation}
or
\begin{equation}
G_b \,=\, \sum_{n=0}^\infty  \frac{1}{m^2 + 2 e F (n + \frac{1}{2})}
\; |n \rangle \langle n| \;.
\end{equation}

\subsection{Complex Proca field}
To calculate the effective action $\Gamma_P$ (for the same gauge field
configuration as before), we make again use of the result for ${\mathcal
  Z}(p,F)$, what in the present case leads to 
\begin{equation}
\Gamma_P (A) \;=\; - \xi \; \int_0^\infty \frac{dT}{T} \, e^{- m^2 T} \;
{\rm Tr}
\Big[ e^{- \, T \, {\hat p}_\alpha {\hat p}_\beta \Gamma^P_{\alpha\beta}}\Big],
\end{equation}
with the same ${\hat p}_\mu$ operators as in the Dirac field case.  The trace
meant both over Hilbert space and Lorentz indices.  In terms of the
annihilation and creation operators ${\hat a}$, ${\hat a}^\dagger$ we have
already introduced, we see that:
\begin{equation}
\Gamma_P (A) \;=\; - \xi \; \int_0^\infty \frac{dT}{T} \, e^{- m^2 T} \;
{\rm Tr} \Big[ e^{- \, \frac{e F}{2} \,T\, {\hat{\mathcal Q}}  }\Big],
\end{equation}
where ${\hat{\mathcal Q}} $ is the operator
\begin{equation}
{\hat{\mathcal Q}} \;=\; \left(\begin{array}{cc}
- ({\hat a} - {\hat a}^\dagger)^2 & i ( {\hat a}^2 - {\hat{ a}}^{\dagger 2}) \\
i ( {\hat{a}}^2 - {\hat{a}}^{\dagger 2})  & ({\hat{a}} + {\hat{a}}^\dagger)^2
\end{array}
\right) \;.
\end{equation}
In order to evaluate the trace, it is convenient to look for the
eigenfunctions and eigenvalues of the ${\mathcal Q}$ operator. We first
rewrite ${\hat{\mathcal Q}}$ as follows:
\begin{equation}
{\hat{\mathcal Q}}\;=\; (2 {\hat{\mathcal N}} + 1) \, I 
\,+\, {\hat a}^2 \,\eta  \,+\, ({\hat a}^\dagger)^2 \, \eta^\dagger \;,
\end{equation}
where $I$ is the $2 × 2$ identity matrix, while $\eta$ denotes the
nilpotent matrix:
\begin{equation}
{\mathbf {\eta}} \;=\; \left(\begin{array}{cc} -1 & i \\ i & 1\end{array} 
\right) \;.
\end{equation}
Eigenvalues $\lambda$ and their corresponding eigenvectors $|\Psi \rangle$ of
${\hat{\mathcal O}}$ may be found, for example, by decomposing (an arbitrary)
$|\Psi \rangle$ as follows:
\begin{equation}
|\Psi \rangle \;=\; |e_+\rangle \otimes |\chi_+ \rangle  \,+\, |e_-\rangle
\otimes |\chi_- \rangle 
\end{equation}
where $|e_±\rangle$ are  two-component vectors:
\begin{equation}
|e_±\rangle \;=\; \frac{1}{\sqrt{2}} \, \left( \begin{array}{c}
 1 \\ ± i
\end{array} 
\right) \;,
\end{equation}
which are obviously linearly independent and satisfy:
\begin{eqnarray}
\eta | e_+ \rangle \;=\; - 2 \, |e_-\rangle  \;\;&,&\;\; 
\eta^\dagger | e_- \rangle \;=\; - 2 \, |e_+\rangle \nonumber\\
\eta^\dagger | e_+ \rangle \;=\; 0 \;\;&,&\;\; \eta | e_- \rangle \;=\; 0 \;, 
\end{eqnarray}
while $|\chi_± \rangle$ are general Hilbert space vectors (scalars with respect
to the Lorentz group).

Inserting the general decomposition into the eigensystem equation, we obtain:
\begin{eqnarray}
\big(2 \, {\hat{\mathcal N}} + 1 \,-\, \lambda \big) \, |\chi_+ \rangle
\;+\; 2 \, ({\hat a}^\dagger)^2 |\chi_- \rangle &=& 0 \nonumber\\
2 \, {\hat a}^2 |\chi_+ \rangle \;+\; \big(2 \, {\hat{\mathcal N}} + 1 \,-\, 
\lambda \big) \, |\chi_- \rangle &=& 0 \;.
\end{eqnarray} 
Now it becomes trivial to solve the last system, for example by using the
basis of eigenstates of the number operator for $|\chi_± \rangle$:
\begin{equation}
|\chi_± \rangle \;=\; \sum_{n=0}^\infty C_n^{(±)} \, | n \rangle \;,
\end{equation}
where $|n\rangle$ denote the eigenvalues of the number operator.  This yields
recurrence relations for the $C_n^{(±)}$'s whose solutions are polynomials
only if $\lambda$ equals an odd integer. Otherwise, the resulting
eigenfunctions are not regular, and must therefore be discarded.  For the
regular solutions, $\lambda = 2 l + 1$, $l = 0,1,\ldots$, there is no
degeneracy. Thus:
\begin{equation}
\Gamma_P (A) \;=\; - \xi \; \int_0^\infty \frac{dT}{T} \, e^{- m^2 T} \;
\sum_{l=0}^\infty  e^{- \, \frac{e F}{2} \,T\, (2 l + 1) } \;,
\end{equation}
which may be integrated over $T$ to obtain:
\begin{equation}
\Gamma_P(A) \;=\; - \xi \;\sum_{l=0}^\infty \ln \big[ m^2 \,+\, \frac{e F}{2}
 ( 2 l + 1) \big]\;.
\end{equation}
Of course, this is equivalent to the massive scalar field, with the trivial
replacement: $e F \to \frac{e F}{2}$: 
\begin{equation}
\Gamma_P (A) \;=\;  \xi  \frac{d}{ds} \Big[ ( e F)^{-s} \, \zeta_H\big(s;
\frac{1}{2} +
\frac{m^2}{e F}\big) \Big] {\Big |}_{s=0} \;.
\end{equation}
\section{Generalization to $d=2k$ dimensions}\label{sec:dqt2}
The calculations of the previous section may be easily generalized to the case
of a constant $F_{\mu\nu}$ field configuration in $d=2k$
dimensions~\footnote{The essential features of the problem are the same for odd
  dimensions but they present subtleties which deserve a separate
  treatment \cite{next}.}.  Indeed, one easily
sees that the effective action for the fermionic case shall be given by an
expression which is formally identical to (\ref{eq:add1})
\begin{equation}
\Gamma_f(A) \,=\, \int_0^\infty \frac{dT}{T} \, e^{- m T} \, \int {\mathcal
D}p \;
{\rm tr}\Big[ {\mathcal P} e^{- i \int_0^T d\tau \not p (\tau)} \Big] \;
{\mathcal Z}(p,F)
\end{equation}
where ${\mathcal Z}(p,F)$ is given by:
\begin{equation}\label{eq:defzpdgt2}
{\mathcal Z}(p,F)\;=\; \int_{x(0)=x(T)}\,  {\mathcal D}x \;\;
e^{i \int_0^T d\tau  {\dot x}_\mu(\tau) p_\mu(\tau) } \;  e^{\frac{i e }{2}
\int_0^T
  d\tau {\dot x}_\mu(\tau) F_{\mu\nu} x_\nu(\tau) }\;,
\end{equation}
as follows from the gauge field configuration:
\begin{equation}
A_\mu (x) \;=\; - \frac{1}{2} \, F_{\mu\nu} x_\nu,
\end{equation}
which satisfies the gauge-fixing condition $\partial\cdot A = 0$.

The easiest way to calculate ${\mathcal Z}(p,F)$ is to reduce the
problem to a set of decoupled $1+1$-dimensional systems, and then to
take advantage of the results of the previous section. That may be
done by using the fact that ${\mathbf F} \equiv (F_{\mu\nu})$ is a real
antisymmetric matrix; hence it may be reduced to a block-diagonal form
${\mathbf f}$ by performing a similarity transformation with an
orthogonal matrix ${\mathbf R}$:
\begin{equation}
{\mathbf F} \;=\; {\mathbf R}^T \, {\mathbf f} \, {\mathbf R} \;.
\end{equation}

Each one of the blocks is $2 × 2$ and antisymmetric, so that the
reduced matrix has the following structure:
\begin{equation}
{\mathbf f} \;=\;
\left(\begin{array}{ccccccc}
   0   &  f^{(1)} & 0 &   0    &  0  & \ldots  & 0 \\
- f^{(1)} &     0  & 0 &   0    &  0 & \ldots  & 0 \\
   0   &     0  & 0 &  f^{(2)}  & 0  & \ldots   & 0 \\
   0   &     0  & - f^{(2)} & 0    & 0  & \ldots & 0 \\
   \ldots  &        &      &      &    &   &   \\
   0   &     0  & 0   &  \ldots      & 0    & 0   &  f^{(k)} \\
   0   &     0  & 0   &  \ldots      & 0    & - f^{(k)} &    0 \\
\end{array} \right) \; ,
\end{equation}
where the $f^{(a)}$, ($a = 1, \ldots , k$) are real numbers, which we
assume to be different from zero (although the particular case of one
or more of them being equal to zero may of course be dealt with at the
end of the calculation).  Then we redefine the momenta and coordinates
in the path integral, according to the following transformation: $p_\mu
\to ({\mathbf R^{-1}})_{\mu\nu} p_\nu$, $x_\mu \to ({\mathbf
R^{-1}})_{\mu\nu} x_\nu$.
The $\gamma$ matrices are also redefined with ${\mathbf R}$ and of course
we arrive to an equivalent representation of the Clifford algebra. We
use the same notation for the new $\gamma$-matrices although we have the
new representation in mind.

The general form of the matrix ${\mathbf F}$ can be further simplified
in some particular cases, when there are some extra restrictions on
the configuration. An interesting example corresponds to $d=4$, where
one has the possibility of considering a self-dual field:
\begin{equation}
{\tilde F}_{\mu\nu} \;=\; F_{\mu\nu} \;\;, \;\;\;\;
{\tilde F}_{\mu\nu} \;\equiv \; \frac{1}{2} \epsilon_{\mu\nu\rho\lambda}
F_{\rho\lambda} \;.
\end{equation}
This relation implies that ${\mathbf F}^2 = - f^2 \, {\mathbf I}$,
where ${\mathbf I}$ is the unit matrix and \mbox{$f^2 = \frac{1}{4}
 F_{\mu\nu} F_{\mu\nu}$}. Then the two blocks in the canonical form
for ${\mathbf F}$ are degenerate:
\begin{equation}
{\mathbf f} \;=\;
\left(\begin{array}{cccc}
   0   &  f & 0 &   0 \\
- f &     0  & 0 &   0  \\
   0   &     0  & 0 &  f  \\
   0   &     0  & - f &  0
\end{array} \right) \; ,
\end{equation}
what does simplify some calculations.

Rather than using the index $\mu$, we introduce the notation $p^{(a)}_i$
($a = 1, \ldots, k$, $i = 1, 2$), which distinguishes the components
according to the $2 × 2$ block they belong to. The same convention is
adopted for $x_\mu$. Then:
\begin{equation}\label{aux1}
\Gamma_f(A) \,=\, \int_0^T \frac{dT}{T} \, e^{- m T} \, \int {\mathcal D}p \;
{\rm tr}\Big[ {\mathcal P} e^{- i \int_0^T d\tau \not p (\tau)} \Big] \;
{\mathcal Z}(p,F) \; ,
\end{equation}
where
$$
{\mathcal Z}(p,F)\;=\; \prod_{a = 1}^k \; \int_{x^{(a)}(0)=x^{(a)}(T)} \,
{\mathcal D}x^{(a)} \; \exp\Big\{ i \int_0^T d\tau \big[ {\dot x}^{(a)}_i
(\tau)
p^{(a)}_i(\tau)
$$
\begin{equation}\label{eq:defzpdgt21}
+ \, i\, \frac{e}{2} f^{(a)} \varepsilon_{ij} {\dot x}^{(a)}_i(\tau)
x^{(a)}_j (\tau) \big] \Big\} \;.
\end{equation}
Of course, for each value of $a$ we have an integral which is
identical to the one for the $1+1$ dimensional case. Thus:
\begin{eqnarray}
{\mathcal Z}(p,F) &=& \prod_{a=1}^k \Big( \xi^{(a)}\;
\lim_{n\to\infty}  \; (\frac{2 \pi}{e f^{(a)}})^{n} \left\{e^{- \frac{i}{e
f^{(a)}} \,  \sum_{k=1}^n p_2^{(k)} (
    p_1^{(k+1)} - p_1^{(k)} )} \right\} \Big) \;,
\end{eqnarray}
a result which we include into (\ref{aux1}), to obtain:
$$
\Gamma_f(A) \;=\; \Big[\prod_{a=1}^k \xi^{(a)} \Big] \, \int_0^\infty
\frac{dT}{T} \,
e^{- m T}
$$
$$
\int_{p_1^{(a)}(0)=p_1^{(a)}(T)} \, \prod_{a=1}^k {\mathcal D} {\hat p}^{(a)}
\, {\rm tr}\Big\{ {\mathcal P} \exp \big[- i \int_0^T d\tau \gamma^{(a)}
p^{(a)}_j(\tau) \big] \Big\}
$$
\begin{equation}\label{eq:gammafn}
× \exp \big[- i \int_0^T d\tau \,  \sum_{a=1}^k \frac{1}{e f^{(a)}} {\dot
  p}^{(a)}_1(\tau) p^{(a)}_2(\tau) \big]\;,
\end{equation}
where:
\begin{equation}
\widehat{{\mathcal D}p^{(a)}}\;=\; \prod_{0 < \tau \leq T}
\frac{dp^{(a)}_1(\tau)
  dp^{(a)}_2(\tau)}{2 \pi e f^{(a)}} \;.
\end{equation}
Of course, the expression for $\Gamma_f(A)$ in (\ref{eq:gammafn}) may be
converted to:
\begin{equation}
\Gamma_f(A) \;=\; \Big[\prod_{a=1}^k \xi^{(a)} \Big] \, \int_0^\infty
\frac{dT}{T} \,
e^{- m T}  \, {\rm Tr}\Big[e^{- i T \sum_{a=1}^k \not {\hat p}^{(a)}} \Big]\;,
\end{equation}
where \mbox{$\not \! {\hat p}^{(a)} \equiv \gamma^{(a)}_1 {\hat p}^{(a)}_1 +
  \gamma^{(a)}_2 {\hat p}^{(a)}_2$} ($a$ is not summed). The ${\hat
  p}^{(a)}$ operators verify the commutation relations:
\begin{equation}
[{\hat p}^{(a)}_j \,,\, {\hat p}^{(b)}_k ]\;=\; - i \, f^{(a)} \,
\delta^{ab} \, \varepsilon_{jk} \;,
\end{equation}
and the trace is over Dirac and Hilbert space.
Since the $\gamma$ matrices satisfy the anticommutation relations:
\begin{equation}
\{ \gamma^{(a)}_j \,,\, \gamma^{(b)}_k \} \;=\;  2 \, \delta^{ab} \,
\delta_{jk} \;,
\end{equation}
we easily see that:
\begin{equation}
\not \! {\hat p}^{(a)} \, \not \! {\hat p}^{(b)}  \;=\; 0 \;\;, \;\;
\forall a \neq b \;.
\end{equation}
Then
$$
\Gamma_f(A) \;=\; \Big[\prod_{a=1}^k \xi^{(a)} \Big] \, \int_0^\infty
\frac{dT}{T} \,
e^{- m T}
$$
\begin{equation}
× \Big[1 \,+\, \sum_{n_1, \ldots ,n_k=1}^\infty \sum_{a=1}^k \big( e ^{- i
T \sqrt{2 e f^{(a)}
    n_a }}
  +  e ^{+ i T \sqrt{2 e f^{(a)} n_a }} \big) \Big]\;,
\end{equation}
or, doing the integral
\begin{equation}
\Gamma_f(A) \;=\; - \, \Big[\prod_{a=1}^k \xi^{(a)} \Big]
× \Big[ \ln (m) \,+\, \sum_{n_1=1, \ldots ,n_k=1}^\infty  \ln \big( m^2 +
\sum_{a=1}^k ( e f^{(a)} n_a)  \big) \Big] \;.
\end{equation}
which upon regularization leads to the known result
(\cite{Blau:1988iz}).

\section{Vortex-like background in $1+1$ dimensions}\label{sec:vortex}
In this section we consider a singular background, corresponding to a
gauge field $A_\mu$ such that:
\begin{equation}
\oint_{\mathcal C} dx_\mu \, A_\mu(x) \;=\; \int_{\mathcal S(C)} \, d^2x \,
\varepsilon_{\mu\nu} \, \partial_\mu A_\nu \;=\; \Phi \,,
\end{equation}
where ${\mathcal S(C)}$ is the planar region enclosed by ${\mathcal
  C}$, and $\Phi$ is a constant. 
In the symmetric gauge:
\begin{equation}
A_\mu(x) \;=\; - \frac{\Phi}{2\pi} \, \varepsilon_{\mu\nu} \,\frac{x_\nu}{x^2} \;. 
\end{equation}

Let us first consider a complex scalar field. In the expression that
leads to its propagator, (\ref{eq:dbprop}), we integrate out the $p$
variable, obtaining:
$$
G_b(x,y) \;=\; \int_0^\infty \, dT\, e^{- m^2 T} \; \int_{x(0)=y}^{x(T)=x}
\,{\mathcal D} x 
$$
\begin{equation}
×\, \exp{ - \int_0^T  d\tau \,\big[\frac{1}{4} {\dot x}^2(\tau)
  \,+\,i \,e\, {\dot x}_\mu(\tau) A_\mu(\tau) \big]}\;,
\end{equation}
where we have absorbed a normalization constant in the integration
measure for $x$.
Next we rescale the time variable: $\tau \;=\; T \, t$, $0\leq t< 1$ to
derive the equivalent representation
$$
G_b(x,y) \;=\; \int_0^\infty \, dT\, e^{- m^2 T} \; \int_{x(0)=y}^{x(1)=x}
\,{\mathcal D} x 
$$
\begin{equation}\label{eq:equiv}
×\, \exp{ - \int_0^1 dt \,\big[\frac{1}{4 T} {\dot x}^2(t)
  \,+\,i \,e\, {\dot x}_\mu(t) A_\mu(t) \big]}\;.
\end{equation}
And we recognize (\ref{eq:equiv}) as the propagator for a
non-relativistic particle of `mass' $(2 T)^{-1}$ moving in two spatial
dimensions in a vortex-like background, corresponding to a time
interval equal to $-i$. Recalling the results of (\cite{gerry}), we
may write an exact expression for the result of the path integral over
$x_\mu(\tau)$:
$$
\int_{x(0)=y}^{x(1)=x}\, e^{- \int_0^1 dt \,\big[\frac{1}{4 T} {\dot x}^2(1)
  \,+\,i \,e\, {\dot x}_\mu(t) A_\mu(t) \big]}\,=\,
\frac{e^{-\frac{1}{4 T} (x^2 + y^2)}}{4 \pi T} \;\sum_{n=-\infty}^{+\infty}
e^{-i e n \Phi} \;
$$
\begin{equation}
× \, \int_{-\infty}^{+\infty} d\lambda \, e^{i \,\lambda\,\big(\theta_y - \theta_x + 2 \pi n \big)} \, I_\lambda
\Big(\frac{|x| |y|}{2 T}\Big) \;,
\end{equation}
where $\theta_x$ and $\theta_y$ are the angular coordinates of $x$ and $y$,
respectively, while $I_\lambda$ is the modified Bessel function of order
$\lambda$.

The sum over $n$ can be transformed by means of Poisson's summation
formula:
\begin{equation}
\sum_{n=-\infty}^{+\infty} e^{i n (2 \pi \lambda - e \Phi)} \;=\; \sum_{k=-\infty}^{+\infty} \delta \big(\lambda
- \frac{e \Phi}{2\pi} +  k \big) \;,
\end{equation}
and this allows us to integrate out $\lambda$. Then: 
$$
G_b(x,y) \;=\; \int_0^\infty \,\frac{dT}{4 \pi T} \, e^{-m^2 T}\,e^{-\frac{1}{4
    T} (x^2 + y^2)} \;×
$$
\begin{equation}
× \sum_{k=-\infty}^{+\infty} e^{i \big[k + \frac{e \Phi}{2 \pi} (\theta_y -\theta_x)\big]} \; I_{k
  + \frac{ e \Phi}{2 \pi}} \Big(\frac{|x| |y|}{2 T}\Big) \;.
\end{equation}

Following~\cite{gerry}, we perform a unitary transformation in order
to absorb the phase factor $e^{i\frac{e \Phi}{2 \pi} (\theta_y -\theta_x)}$ into the
propagator. At the level of fields, the transformation is
\begin{equation}
\phi(x) \;=\; e^{ - i\frac{e \Phi}{2\pi} \, \theta_x} \; {\tilde \phi}(x)\;,\;\;\;
{\bar \phi}(y) \;=\; e^{ i \frac{e \Phi}{2\pi} \, \theta_y} \; {\bar {\tilde \phi}}(y)\;,
\end{equation} 
while for the propagator we have:
$$
{\tilde G}_b(x,y) \;\equiv\; \langle {\tilde \phi}(x) \, {\bar{\tilde \phi}}(y)\rangle \;=\; 
 e^{ i \frac{e \Phi}{2\pi} (\theta_x - \theta_y) } \; G_b(x,y)
$$
\begin{equation}
=\;  \int_0^\infty \,\frac{dT}{4 \pi T} \, e^{-m^2 T}\,e^{-\frac{(x^2 + y^2)}{4
    T}}
× \sum_{k=-\infty}^{+\infty} e^{i k (\theta_y -\theta_x)} \; I_{k + \frac{ e \Phi}{2 \pi}} \Big(\frac{|x| |y|}{2 T}\Big) \;,
\end{equation}
and, of course, physical observables are not sensitive to this
transformation.

Then we see that the form of the propagator ${\tilde G}_b(x,y)$ is
periodic in the difference between the angles $\theta_y-\theta_x$, and it has
the form of an expansion in `partial waves':
\begin{equation}
{\tilde G}_b (x,y) \;=\; \sum_{k=-\infty}^{+\infty} \,e^{i k (\theta_y -\theta_x)} \,  
{\tilde G}_b^{(k)} (|x|,|y|)
\end{equation}
where each ${\tilde G}_b^{(k)}$ corresponds to the $k^{th}$ partial
wave:
\begin{equation}\label{eq:partial1}
{\tilde G}_b^{(k)}(|x|,|y|) \;=\; 
\int_0^\infty \,\frac{dT}{4 \pi T} \, e^{-m^2 T}\,e^{-\frac{(x^2 + y^2)}{4
    T}} I_{k + \frac{ e \Phi}{2 \pi}} \Big(\frac{|x| |y|}{2 T}\Big) \;.
\end{equation}

The integral over $T$ cannot be performed exactly, but we may instead
integrate term by term by expanding the Bessel function in a
`long-distance' approximation: $|x| |y| >> T$. This amounts to large
values of $|z|$ in the following asymptotic expansion
\begin{eqnarray}\label{eq:besselexp}
I_\nu (z) &\sim&\frac{e^z}{(2\pi z)^{\frac{1}{2}}} \; \sum_{l=0}^{+\infty} \,
  \frac{(-1)^l}{(2 z)^l} \; 
\frac{\Gamma(\nu + l +\frac{1}{2})}{l!\, \Gamma(\nu - l +\frac{1}{2})} \nonumber\\
&+&
\frac{e^{-z+(\nu+1/2) i \pi}}{(2\pi z)^{\frac{1}{2}}} \; \sum_{l=0}^{+\infty} \,
  \frac{1}{(2 z)^l} \; 
\frac{\Gamma(\nu + l +\frac{1}{2})}{l!\, \Gamma(\nu - l +\frac{1}{2})} \;.
\end{eqnarray}
Inserting the first line in the previous expansion (the second is
exponentially suppressed) for for each partial wave in
(\ref{eq:partial1}),
and integrating out $T$, we obtain:
$$
{\tilde G}_b^{(k)}(|x|,|y|) \;=\; \frac{1}{2 \pi^{\frac{3}{2}}} \, 
\sum_{l=0}^{+\infty} \; \frac{(-1)^l}{l !} \; 
\frac{\Gamma(k + b +\frac{1}{2}+l)}{\Gamma(k + b +\frac{1}{2}-l)}
$$
\begin{equation}\label{eq:partial2}
\frac{1}{(|x| |y|)^{l+1/2}} \; \big(\frac{\big| |x| - |y| \big|}{2 m}\big)^{l +
    \frac{1}{2}}\;
K_{l + \frac{1}{2}}\big(m \big| |x|-|y| \big|\big) \;, 
\end{equation}
where we introduced the dimensionless flux: $b \equiv \frac{e \Phi}{2\pi}$.
\section{Exact results for the massless Dirac field in $1+1$
  dimensions}\label{sec:exact}
 We will consider now the massless
fermionic propagator in 1+1 dimension in the presence of an arbitrary
external field. The result is known to be solvable \cite{XXX}.  We
 start with the fermionic propagator given by (\ref{eq:dprop})
%
with the vector field $A_\mu$ written in the form 
\begin{equation}
A_\mu(x) = \partial_\mu \chi + i \epsilon_{\mu \nu} \partial_\nu \phi. 
\end{equation}
this is always possible in 1+1 dimension.
The last factor in the path integral in Eq. (\ref{eq:dprop}) reads
\begin{equation}
-i e \int_0^T d\tau {\dot x}(\tau)\cdot A[x(\tau)] = -i e \int_0^T d\tau
 {\dot x_\mu}(\tau)\cdot (\partial_\mu \chi + i \epsilon_{\mu \nu} 
\partial_\nu \phi)
\end{equation}
The first term gives a total derivative in the integral over the path which is
independent of the path and can be taken out of the path integral
\begin{equation}
-i e \int_0^T d\tau {\dot x}(\tau)\cdot A[x(\tau)] = -i e (\chi(x) -\chi(y))
-i e \int_0^T d\tau  {\dot x_\mu}(\tau) i \epsilon_{\mu \nu} \partial_\nu \phi
\end{equation}
As was shown in~\cite{migdal,Fosco} terms with a $\dot x(\tau)$ can be
substituted by a $\gamma$ inside the path integral. Thus the path integral
can be written in the following form
\begin{eqnarray}
G_f(x,y,m) &=& e^{-i e  (\chi(x) -\chi(y)) } \int_0^\infty dT \,e ^{- m T} \, 
\int_{x(0) = y}^{x(T) = x} {\mathcal D}p
{\mathcal D}x \; e^{i \int_0^T d\tau  p(\tau) \cdot{\dot x}(\tau) } \nonumber\\
&×& {\mathcal P}[ e^{- i \int_0^T d\tau {\not p}(\tau) - e \gamma_5
\not \partial \phi}] \;,
\end{eqnarray}
where we have used the fact that, in $2$ dimensions,
\begin{equation}
i \gamma_\mu \epsilon_{\mu \nu} \partial_\nu = -\gamma_5 \not \! \partial \;.
\end{equation}

Consider now the action of the following infinitesimal transformation
on a free fermionic propagator
\begin{equation}
G^{\alpha}_f(x,y,m) = e^{ie \gamma_5 \alpha(x)} G_f(x,y,m) e^{ie \gamma_5 \alpha(y)}
\end{equation}
where $\alpha(x)$ is infinitesimal. Then the first order variation can be
written as follows
\begin{eqnarray} 
\delta G^{\alpha}_f(x,y,m) & = & G^{\alpha}_f(x,y,m) -G^0_f(x,y,m) \\ \nonumber
& = & 
i e \gamma_5 \alpha(x) G_f(x,y,m) + i e G_f(x,y,m) \gamma_5 \alpha(y) \\ \nonumber
& = & i e \gamma_5 \alpha(x)  G_f(x,y,m) -i e \gamma_5 \alpha(y) G_f(x,y,-m).
\nonumber 
\end{eqnarray}
{From} the last equality it follows that for massless Dirac fields we have
\begin{equation}
\delta G^{\alpha}_f(x,y,m=0) = i e \gamma_5 (\alpha(x) -\alpha(y) )  
G_f(x,y,m=0).
\end{equation}
The same will apply for the propagator in the presence of an external
electromagnetic field, for the propagator in (\ref{eq:dprop_phi}). Then we
will have
\begin{eqnarray}
G^{\alpha}_f(x,y,\phi) & = & [1+ i e \gamma_5 (\alpha(x) -\alpha(y) ]
G_f(x,y,\phi) \\ \nonumber
& = & e^{i e \gamma_5 [\alpha(x) -\alpha(y)]} G_f(x,y,\phi) \nonumber
\end{eqnarray}
The factor can be introduced in the path integral, it appears as a term
of the same form of the $\phi$ 
\begin{eqnarray} \label{eq:dprop_phi}
G_f(x,y,m) &=& e^{-i e  (\chi(x) -\chi(y)) } \int_0^\infty dT \,e ^{- m T} \, 
\int_{x(0) = y}^{x(T) = x} {\mathcal D}p
{\mathcal D}x \; e^{i \int_0^T d\tau  p(\tau) \cdot{\dot x}(\tau) } \nonumber\\
&×& {\mathcal P}[ e^{- i \int_0^T d\tau {\not p}(\tau) - e \gamma_5
\not \partial (\phi+\alpha)}] \;.
\end{eqnarray}
One may then choose $\alpha$ to cancel the $\phi$ term. The final result
is therefore:
\begin{equation}
G_f(x,y,\phi) = e^{-ie \gamma_5 \phi(x)} G_f(x,y) e^{-ie \gamma_5 \phi(y)} \;,
\end{equation}
an expression could be useful in the numerical implementation of
fermionic path integrals, which is a long-standing problem.

\section{Anomalies and path roughness}\label{sec:anomaly}
The anomaly ${\mathcal A}$ corresponding to a global axial
transformation of the Dirac fields:
\begin{eqnarray}\label{eq:axial}
\delta\psi &=& i \, \xi \, \gamma_5 \, \psi \nonumber\\
\delta {\bar\psi} &=& i \, \xi \, {\bar \psi} \gamma_5 \;,
\end{eqnarray}
is determined by the (functional and Dirac) regularized trace of $\gamma_5$,
\begin{equation}
{\cal A} \;=\; \lim_{M \to \infty}  {\rm Tr} \Big[ \gamma_5 \, f(\frac{-\not \!\!D^2}{M^2}) \Big]
\end{equation}  
where $f$ is a regulating function which verifies the conditions:
\begin{equation}
f(0)\,=\,1 \;\;,\;\;\; f(±\infty)\,=\,  f'(±\infty) \,=\, f''(±\infty)\,=\ldots = 0 \;.
\end{equation}

With the choice $f(x) = 1/(1+ x^2)$, and after a little algebra we
obtain the equivalent expression:
\begin{equation}
{\mathcal A} \;=\; \lim_{M \to \infty}  M \, {\rm Tr} \Big[ \gamma_5 \, \big( \not
\!\! D +
M \big)^{-1}  \Big] \;.
\end{equation}
This of course may be written in terms of the propagator for a massive
field, as follows:
\begin{equation}
{\mathcal A} \;=\; \lim_{M \to \infty}  M \, {\rm tr} \, \Big[ \gamma_5 \, G_f (x,x) \Big]
\end{equation}
where ${\rm tr}$ denotes the trace over spin indices only.
Introducing the worldline representation, we see that
\begin{eqnarray}
{\mathcal A} &=& \lim_{M \to \infty}  M \, {\rm tr} \, \Big[ \gamma_5 \, G_f (x,x) \Big]
\nonumber\\
&=& \int_0^\infty dT \,e ^{- m T} \, \int_{x(0) = x(T)} {\mathcal D}p
{\mathcal D}x \; e^{i \int_0^T d\tau  p(\tau) \cdot{\dot x}(\tau)} \nonumber\\ 
&× & {\rm tr} \Big[\gamma_5 {\mathcal P} e^{- i \int_0^T d\tau {\not p}(\tau) } \Big]\; 
e^{- i e \int_0^T d\tau {\dot x}(\tau)\cdot A[x(\tau)]} \;.
\end{eqnarray}
Then we note that $\gamma_5$ can, {\em inside the path integral}, be
written purely in terms of ${\dot x}_\mu(\tau)$ and its derivatives, by
using relations of the kind already introduced in~\cite{Fosco}. There
we related the average of $\gamma_\mu$ to $\dot{x}_\mu$. Now, with a precise
form that shall depend on the number of spacetime dimensions, we can
relate the average of $\gamma_5$ to (the average of) a product of
$\dot{x}_\mu$ and its derivatives. Since the order in the product
matters, we have to introduce a time splitting between the different
factors.

To be more concrete, let us consider the case $d=2$. Since $\gamma_5 = -i \gamma_1
\gamma_2$,
\begin{eqnarray}
{\mathcal A}_{d=2} &=& -i \, \lim_{M \to \infty}  M \; \int_0^\infty dT \,e ^{- m T} \, 
\int_{x(0) = x(T)} {\mathcal D}p {\mathcal D}x \; e^{i \int_0^T d\tau  p(\tau)
  \cdot{\dot x}(\tau)} \nonumber\\ 
&× & {\dot x}_1 (T) {\dot x}_2 (T-\epsilon)  
{\rm tr}\Big[{\mathcal P} e^{- i \int_0^T d\tau {\not p}(\tau) } \Big]\; 
e^{- i e \int_0^T d\tau {\dot x}(\tau)\cdot A[x(\tau)]} \;,
\end{eqnarray}
where $\epsilon$ is a positive infinitesimal parameter, to be taken to zero
at the end of the calculation. Of course, we may use the more
symmetric expression:
\begin{eqnarray}
{\mathcal A}_{d=2} &=& - \frac{i}{2} \, \lim_{M \to \infty}  M \; \int_0^\infty dT \,e ^{- m T} \, 
\int_{x(0) = x(T)} {\mathcal D}p {\mathcal D}x \; e^{i \int_0^T d\tau  p(\tau)
  \cdot{\dot x}(\tau)} \nonumber\\ 
&× &  \varepsilon_{\mu \nu} {\dot x}_\mu (T) {\dot x}_\nu (T-\epsilon)  
{\rm tr}\Big[{\mathcal P} e^{- i \int_0^T d\tau {\not p}(\tau) } \Big]\; 
e^{- i e \int_0^T d\tau {\dot x}(\tau)\cdot A[x(\tau)]} \;,
\end{eqnarray}
for which we use the notation:
\begin{equation}
{\mathcal A}_{d=2} \;=\; - \frac{i}{2} \, \lim_{M \to \infty}  M \; \int_0^\infty dT \,e ^{- m T} \, 
\langle \varepsilon_{\mu \nu} {\dot x}_\mu (T) {\dot x}_\nu (T-\epsilon) \rangle  \;.
\end{equation}
Expanding for small $\epsilon$,
\begin{equation}
{\mathcal A}_{d=2} \;=\; \frac{i}{2} \, \lim_{M \to \infty}  M \; \int_0^\infty dT \,e ^{- m T} \, 
\epsilon \; \langle \varepsilon_{\mu \nu} {\dot x}_\mu (T) {\ddot x}_\nu (T) \rangle  \;,
\end{equation}
which shows that, in order to have a non-vanishing chiral anomaly in
$d=2$, the correlation function between $\dot{x}_\mu$ and $\ddot{x}_\nu$,
has to be singular at the coincidence limit   \footnote{Similar
qualitative results have been obtained in reference \cite{Hammerling}
for the effective action in the second order formalism.}. More
precisely, the correlator between the velocity and the acceleration in
orthogonal directions has to diverge as $\epsilon^{-1}$. The object that
appears inside the average symbol can also be given a geometric
interpretation: indeed, if one uses the natural parameterization for the paths:
$x_\mu = x_\mu(s)$ ($\Rightarrow \;\dot{x}^2 = 1$), then:
\begin{equation} 
\frac{1}{2}\,\langle \varepsilon_{\mu \nu} {\dot x}_\mu (T) {\ddot x}_\nu (T) \rangle  \;=\;
\langle \frac{d\theta(s)}{ds} \rangle 
\end{equation}
where $\theta(s)$ is the angle between the tangent vector to the curve and
a (fixed) reference line~\footnote{This object also appears in the
`Polyakov phase factor'. See, for example~\cite{polyakov}.}.

In this way, we can rephrase our statement about the roughness:  
$\langle \frac{d\theta(s)}{ds}\rangle$ is singular ($\sim\epsilon^{-1}$) when the anomaly is
non-vanishing.

In $d=4$, an entirely similar procedure yields
 \begin{equation}
{\mathcal A}_{d=4} = - \frac{i}{6} \lim_{M \to \infty}  M \int_0^\infty dT e ^{- m T} \, 
\langle \varepsilon_{\mu \nu\rho\sigma} {\dot x}_\mu (T) {\dot x}_\nu (T-\epsilon) {\dot x}_\rho (T-2\epsilon) {\dot
  x}_\sigma (T-3\epsilon) \rangle  \,,
\end{equation}
which, in order for the lhs to be different from zero, requires a
different kind of singularity in the correlation functions, since it
involves up to to de fourth derivative of the curve. Its geometrical
meaning is less appealing than in four dimensions (it is proportional
to the rate of variation of the bi-torsion of the curve).
\section{Conclusions}\label{sec:concl}
We have carried further the first-order spin formalism for the
worldline, providing new tests and applications for Migdal's
construction.  We have thus obtained with this method, new expressions
for the propagators and effective actions of the Dirac, Proca and
complex scalar fields coupled to Abelian gauge fields.  A remarkable
result is the universality of the path order spin factor.  In fact
this can be seen as a natural consequence of the geometric
representation and are extendible to higher spin fields.

For constant electromagnetic fields in two dimensions we have shown
that our results agree with the results from the zeta-function
renormalization.  The results have been then generalized to $d$ even
dimensions.  Notice that the first order formalism one can incorporate
the quantum fluctuations of the gauge field.  We have obtained an
exact expression for the fermionic propagator in $1+1$ dimensions. Other
new results are the vortex configuration and the axial anomaly where
the results can be understood in terms of a very attractive geometric
feature of the contributing paths.

This can provide a useful contribution to the progress in
nonperturbative quantum field dynamics within the worldline. Specially
for its numerical implementation, which appears as a powerful new
alternative~\cite{nos}. Work in progress in odd dimensions indicates
promising features of transferring internal degrees of freedom to
geometrical properties of space time which could hopefully allow to
include non Abelian fields.

\section*{Acknowledgements}
J.S.-G. and R.A.V. thank MCyT (Spain) and FEDER (FPA2005-01963), and
Incentivos from Xunta de Galicia. C.D.F. has been supported by
CONICET.


\end{document}